\documentclass[showpacs,nofootinbib,amsmath,amssymb,amsfonts,english]{revtex4}

\usepackage{babel}

\usepackage{MALC}

\begin{document}

\title{Melvin universe as a limit of the C-metric}

\author{Lenka Havrdov\'a}
\email{lenkahavrdova@seznam.cz}

\author{Pavel Krtou\v{s}}
\email{Pavel.Krtous@mff.cuni.cz}

\affiliation{
  Institute of Theoretical Physics,
  Faculty of Mathematics and Physics, Charles University,\\
  V Hole\v{s}ovi\v{c}k\'{a}ch 2, 180 00 Prague 8, Czech Republic
  }

\date{September 26, 2006}  

\begin{abstract}
It is demonstrated that the Melvin universe representing
the spacetime with a strong \vague{homogeneous} electric field can
by obtained from the spacetime of two accelerated charged 
black holes by a suitable limiting procedure.
The behavior of various invariantly defined geometrical 
quantities in this limit is also studied.
\end{abstract}



\maketitle


\section{Introduction}
\label{sc:intro}

It is a simple exercise to show that the homogeneous electric field in Minkowski
spacetime can be obtained from the Born solution representing test fields of two
uniformly accelerated charged particles by enlarging the distance between these 
particles accompanied with increasing the charges.

One would thus expect that a similar limit of a \vague{homogeneous} field could be obtained
also for a fully gravitationally interacting  electromagnetic field and strong sources.
The non-test analogue of the Born solution is the charged C-metric representing
two uniformly accelerated charged black holes \cite{LeviCivita:1917,KinnersleyWalker:1970}
(cf.~\cite{PravdaPravdova:2000} for review). The gravitationally interacting \vague{homogeneous} 
electric field is described by the Melvin universe
\cite{Melvin:1964,Melvin:1965,Thorne:1965}. 
However, because such a field is infinitely extended, 
the gravitational interaction strongly deforms the spacetime and nontrivially 
changes its asymptotical structure. The limit of the finite sources is thus not completely
straightforward. Most of the na\"\i ve limits of the C-metric lead to 
degenerate results or to empty Minkowski spacetime \cite{Havrdova:2006}.

Let us note that the homogeneous field in Minkowski space can be 
also obtained  by different limiting procedures, some of which have a
strong field analogy, see e.g.\ \cite{EmparanGutperle:2001,GibbonsHerdeiro:2001}.
Our work complements this effort and demonstrates that 
the limit of the distant strong charged accelerated sources
exists, however it is not a trivial generalization of the test case.    

In the next section we present a suitable way to scale 
the parameters and coordinates of the C-metric to obtain the desired limit.
In Sec.~\ref{sc:Melvin} it is shown that the metric obtained indeed 
corresponds to the Melvin universe, and in Sec.~\ref{sc:geomquant}
the behavior of various physical and geometrical quantities
is studied.

\section{Scaling the C-metric}
\label{sc:Cmetric}

Following \cite{HongTeo:2003} we write the C-metric in the form 
\begin{equation}\label{Cmtrc}
  \mtrc= \frac1{\accl^2(x+y)^2}\Bigl(-F\grad t^2+\frac1F \grad y^2 + \frac1G \grad x^2 + G\grad\phi^2\Bigr)\commae
\end{equation}
with ${t\in\realn}$, ${x\in(-1,1)}$, ${y<-x}$, ${\phi\in(-\conpar\pi,\conpar\pi)}$, and 
\begin{equation}\label{FGdef}
\begin{aligned}
  F&=(y^2-1)(1-2\mass\accl y+\charge^2\accl^2 y^2) \commae\\
  G&=(1-x^2)(1+2\mass\accl x+\charge^2\accl^2 x^2) \period
\end{aligned}
\end{equation}
The four roots ${\yc,\ya,\yo,\yi}$ of the function ${F}$ are 
\begin{equation}\label{roots}
  \yc=-1\comma\ya=1\comma
  \yo=\frac1{\accl \charge^2}(\mass-\sqrt{\mass^2-\charge^2})\comma
  \yi=\frac1{\accl \charge^2}(\mass+\sqrt{\mass^2-\charge^2})\period
\end{equation}
 
The metric solves the Maxwell-Einstein equations with electromagnetic field given by
\begin{equation}\label{EMfield}
  \EMF=\charge\,\grad y\wedge\grad t\period
\end{equation}

The C-metric \eqref{Cmtrc} describes two uniformly accelerated charged black 
holes moving in opposite direction along the symmetry axis. 
The constants ${\accl,\mass,\charge}$, and ${\conpar}$
parametrize acceleration, mass, charge and conicity of the symmetry axis. 
The coordinate ${t}$ represents time of static observers around the black holes.
Near the holes, ${y}$ is an (inverse) radial coordinate, which far away from the holes has a bi-spherical character 
(cf.~\cite{GriffithsKrtousPodolsky:2006}). The coordinate ${x}$ is a (cosine of) angle measured
from the symmetry axis, and ${\phi}$ is the rotational angle around the axis. 
The black holes are causally separated by the acceleration horizon ${y=1}$, the outer and inner black hole horizons are
given by ${y=\yo}$ and ${y=\yi}$, respectively.
The conformal infinity is given by the condition ${y=-x}$.
The properties of C-metric were studied, e.g., in 
\cite{KinnersleyWalker:1970,AshtekarDray:1981,Dray:1982,FarhooshZimmerman:1979,Bicak:Bonnor,PravdaPravdova:2000,LetelierOliveira:2001},
the global structure was thoroughly discussed
and visualised recently in \cite{GriffithsKrtousPodolsky:2006}.

To obtain the desired limit of a \vague{homogeneous} field we have to specify the behavior of
the metric parameters ${\accl,\mass,\charge,\conpar}$, and of the coordinates. We fix the 
acceleration parameter ${\accl}$ and the conicity parameter ${\conpar}$ unchanged during the limit. 
The behavior of ${\mass}$ and ${\charge}$ is prescribed implicitly in terms of the roots \eqref{roots} as
\begin{equation}\label{yoilim}
  \yo=1+\tyo\eps\comma \yi=1+\tyi\eps\commae
\end{equation}
where ${\tyo<\tyi}$ are constants parametrizing the limiting procedure,
and ${\eps}$ is a small parameter which will be sent to zero in the limit.
Since the coefficients of the metric functions \eqref{FGdef} are related,
the behavior of the roots of ${F}$ determines also the behavior of
the roots of ${G}$.

Simultaneously we introduce rescaled coordinates ${\{\tau,\y,\x,\ph\}}$ 
\begin{equation}\label{scaledcoor}
  t=\frac1{\tyo\tyi}\,\frac1{\eps^2}\;\tau
     \comma
  y=1+\tyo\tyi\;\eps^2\;\y
     \comma
  x=\xi
     \comma
  \phi=\frac1{1\!\!+\!2\mass\accl\!+\!\charge^2\accl^2}\;\ph
     \period
\end{equation}
If we redefine the conicity parameter
${\conicity=\conpar(1+2\mass\accl+\charge^2\accl^2)}$, the range 
of coordinate ${\ph}$ is ${\ph\in(-\conicity\pi,\conicity\pi)}$.
The coordinate ${\ph}$ is rescaled in such a way that ${c}$ 
gives the actual conicity of the axis between the black holes.
The axis is regular if ${c=1}$.

According to \eqref{scaledcoor}, the coordinate ${y}$ is rescaled by the factor ${\sim\eps^2}$ around the value ${y=1}$.
The limit is thus done in the domain close to the acceleration horizon.
The values ${\yo}$ and ${\yi}$ are rescaled only as ${\sim\eps}$ (cf.\ eq.~\eqref{yoilim})
and thus the corresponding values ${\y_\ohor,\,\y_\ihor}$ of the rescaled coordinate ${\y}$ 
behave as ${\sim\eps^{-1}}$. Therefore, the horizons ${\y_\ohor,\,\y_\ihor}$ disappear in the limit ${\eps\to0}$. 
This corresponds to the intuition gathered in the limit of the Born solution mentioned in the Introduction: the homogeneous electric field should be obtained in the middle between very distant sources. The fact that black holes are indeed \vague{pushed} away in the limit is also confirmed
by the evaluation of their physical distance in Eq.~\eqref{dist} below.

The coordinate ${x}$ is not scaled and the upper bound on ${\y}$ corresponding
to the condition ${y<-x}$ thus also becomes trivial, ${\y<\infty}$, in the limit. 
Finally, the three roots of the function ${G}$ 
degenerate to the value ${x=-1}$ as ${\eps\to0}$.

After the transformations \eqref{yoilim} and \eqref{scaledcoor} 
and sending ${\eps\to0}$ the metric \eqref{Cmtrc} and 
the electromagnetic field \eqref{EMfield} become\footnote{Notice that ${\mass\to1/\accl}$, ${\charge\to1/\accl}$, and ${E\to4\accl}$ in the limit, cf.\ Eqs.~\eqref{melim} below.}
\begin{equation}  \label{Cmtrcresc}
  \mtrc
  =\frac1{\accl^2(1\!+\!\x)^2}\Bigl(-2\y\,\grad\tau^2\!+\!
        \frac1{2\y}\,\grad\y^2\!+\!\frac1{(1\!-\!\x)(1\!+\!\x)^3}\,\grad\x^2\Bigr)
        +\frac1{16\accl^2}(1\!-\!\x^2)\,\grad\ph^2\commae
\end{equation}
and
\begin{equation}  \label{EMfield resc}
  \EMF=\charge\,\grad\y\wedge\grad\tau=\field\frac{(1+\x)^2}{4}\,\tf{\y}\!\wedge\tf{\tau}\period
\end{equation}
Here ${\tf{\y}}$ and ${\tf{\tau}}$ are normalized 1-forms proportional to ${\grad\y}$, and ${\grad\tau}$
and ${E=4\charge\accl^2}$ is the value of the electromagnetic field at the bifurcation 
${\y=0}$, ${\x=1}$ of the acceleration horizon.

\section{Melvin Universe}
\label{sc:Melvin}

The metric \eqref{Cmtrcresc} still inherits the boost-rotational structure of the C-metric in the $\tau$-$\y$ plane. To demonstrate that it describes the Melvin universe, we have to
transform it into coordinates which exhibit more properly the symmetries of the Melvin universe. 
Therefore we perform an additional coordinate transformations introducing the Melvin coordinates\footnote{%
The Melvin coordinate ${\tM}$ is different from the C-metric coordinate~${t}$ used in \eqref{Cmtrc}.}
${\{\tM,\zM,\rho,\ph\}}$:
\begin{equation}  \label{Rindler}
  \y=2\accl^2(-\tM^2+\zM^2)\comma \tanh^{\sign\y}\tau=\frac\tM\zM
\end{equation}
and
\begin{equation}\label{xtorho}
  \x=\frac{1-\bigl(\frac\field2\rho\bigr)^2}{1+\bigl(\frac\field2\rho\bigr)^2}
  \quad\Leftrightarrow\quad
  \Bigl(\frac\field2\rho\Bigr)^2 = \frac{1-\x}{1+\x}\period
\end{equation}
Let us observe that the transformation \eqref{Rindler} is essentially the
transformation from Rindler-like coordinates ${\tau}$ and ${\zeta\propto\sqrt{\abs{\y}}}$
to Minkowski-like coordinates ${\tM}$ and ${\zM}$. It is the transformation from the frame of
uniformly accelerated observers, which is natural for the C-metric, into the frame of globally static observers more natural for the Melvin universe. The coordinate ${\rho}$ could be also given as ${\frac\field2\rho=\tan\frac\tht2}$, where ${\x=\cos\tht}$.

The transformations \eqref{Rindler} and \eqref{xtorho} lead to
\begin{gather}  
  \mtrc=\!\Bigl(1\!+\!\frac{\field^2}{4}\rho^2\Bigr)^2\Bigl(-\grad\tM^2\!+\grad\zM^2\!+\grad\rho^2\Bigr)
       +\Bigl(1\!+\!\frac{\field^2}{4}\rho^2\Bigr)^{\!-2}\rho^2\,\grad\ph^2\commae\label{Melving}\\[2ex]
  \EMF=\field\,\grad\zM\wedge\grad\tM = \frac{\field}{\Bigl(1+\frac{\field^2}{4}\rho^2\Bigr)^2}\,\tf{\zM}\!\wedge\tf{\tM}\commae\label{MelvinF}
\end{gather}
with the coordinate ranges ${\tM,\zM\in\realn}$, ${\rho\in\realn^+}$, and ${\ph\in(-\conicity\pi,\conicity\pi)}$.
These are respectively the metric and electromagnetic field of the Melvin universe \cite{Melvin:1964,Melvin:1965,Thorne:1965}
representing the strong \vague{homogenous} electric field oriented along the symmetry axis.
The field is \vague{homogeneous} in the sense that it does not change 
under translations along the direction of the field.
Indeed, the metric and the field are invariant under the action of 
the static, translational, and rotational Killing vectors 
${\cvil{\tM}}$, ${\cvil{\zM}}$, and ${\cvil{\ph}}$, respectively.
However, the field changes in the direction orthogonal to the symmetry axis.  
The constant ${\field}$ measures the electromagnetic field 
on the axis and ${\conicity}$ the conicity of the axis.

\section{Geometrical quantities}
\label{sc:geomquant}

As a consequence of the expressions \eqref{yoilim} we obtain the following behavior 
of the metric parameters:
\begin{equation}  \label{melim}
\begin{gathered}
  \mass^2\accl^2 = 1-(\tyo+\tyi)\,\eps+\OO(\eps^2)\comma
  \charge^2\accl^2 = 1-(\tyo+\tyi)\,\eps+\OO(\eps^2)\commae\\
  \accl\sqrt{\mass^2-\charge^2} \approx \frac12(\tyi-\tyo)\,\eps\comma
  \field\approx4\accl\comma \conpar\approx\frac\conicity4\period
\end{gathered}
\end{equation}
However, it is more interesting to evaluate geometrically invariant quantities.
The area and surface gravity\footnote{%
The surface gravity ${\surfgr}$ depends on the normalization of the Killing vector.
Here we use the Killing vector ${\accl\cvil{t}}$.}
of the outer black hole horizon are given by
\begin{equation}  \label{Sandk}
\begin{gathered}
  \area_\ohor = \frac{4\pi\conpar}{\accl^2}\frac1{\yo^{\!2}\!-\!1}\approx
     \frac{\pi\conicity}{2\accl^2\tyo}\frac1\eps\commae\\
  \surfgr_\ohor = \frac12\accl\,(\yo^{\!2}\!-\!1)\Bigl(\frac1\yo-\frac1\yi\Bigr)\approx
     \tyo(\tyi-\tyo)\,\eps^2\period
\end{gathered}
\end{equation}
The surface gravity of the acceleration horizon is
\begin{equation}  \label{surfgra}
\surfgr_\ahor = \accl(1-2\mass\accl+\charge^2\accl^2)\approx\accl(2\tyo^2+2\tyi^2+\tyo\tyi)\,\eps^2\period
\end{equation}
Total charge evaluated using the Gauss law is simply ${\totcharge=\conpar\charge\approx\conicity/\field}$.
As we already mentioned, the conicity of the axis ${x=1}$ between the black holes
is given by the parameter ${\conicity}$. It transforms to the conicity of the axis of the Melvin universe as ${\eps\to0}$. 
The conicity of the outer axes, which join the black holes with infinity,
vanishes in the limit. Simultaneously, these axes are pushed away to infinity. 
Finally, the distance of the black hole from the acceleration horizon 
measured along the axis (half of the distance between the holes) is
\begin{equation}  \label{dist}
\distao=\frac1\accl\int_1^\yo\!\!\frac1{(1+y)\sqrt{F}}\;dy\approx
\frac1{\accl\sqrt{2\tyi}}\;\mathrm{K}\Bigl(\frac\tyi\tyo\Bigr)\frac1{\sqrt{\eps}}\commae
\end{equation}
where ${\mathrm{K}(w)}$ is the complete elliptic integral of the first kind.

We see that as ${\eps\to0}$ the black holes are pushed far away from each other, the surface of the outer horizons
increases to infinity, the surface gravity of both the acceleration and the outer horizons vanish, and the
total charge remains finite.

\section{Summary}
\label{sc:summary}

We have shown that the fully gravitationally interacting \vague{homogeneous}  
electric field can be obtained as the limit of the spacetime with localized 
uniformly accelerated sources. Similarly to the limit of the Born solution 
(mentioned in the Introduction) the distance between the sources increases 
to infinity. 

However, unlike the Born solution limit, the total
charge of the sources remains finite. This is related to the fact 
that strong \vague{homogeneous} electric field is gravitationally 
bounded to the vicinity of the symmetry axis and decays as ${\rho\to\infty}$,
cf.\ Eq.~\eqref{MelvinF}. The total electric flux through 
the plane orthogonal to the axis is thus finite.

A more delicate question is the behavior of the mass of the black holes.\footnote{%
It is relatively easy to identify the \vague{net force} ${\mass\accl}$ acting on the accelerated black hole 
in terms of invariantly defined quantities (e.g., taking a difference of tensions---proportional to the conicities---of the strings on the both side of black holes). However, it is not a simple task to invariantly distinguish the physical mass 
and the physical acceleration of the holes. The parameters ${\mass}$ and ${\accl}$ used in the metric \eqref{Cmtrc} are conventional, they depend on a particular choice of the form of the C-metric.} 
Relating the black hole mass to the area of the outer black hole horizon, 
we have shown that it grows to infinity. 

Notice also that the resulting field \eqref{Melving}, \eqref{MelvinF} does not depend on details 
of the limiting procedure---the factors ${\tyo}$ and ${\tyi}$ introduced in \eqref{yoilim}, 
which parametrize how the limit is achieved, disappear completely at the end.

A similar limit can also be performed in the presence of a negative cosmological
constant \cite{Havrdova:2006}. Rescaling the C-metric with ${\Lambda<0}$ and with 
an over-critical acceleration (cf.~\cite{Krtous:2005}) leads to a spacetime with 
an electromagnetic string in the anti-de~Sitter universe \cite{DiasLemos:2002}.

\begin{acknowledgments}
This work was supported in part by the grant GA\v{C}R 202/06/0041.
\end{acknowledgments}


\end{document}